\documentstyle[12pt]{article}
\input epsf
\title{Hydrodynamics of a Bose condensate: beyond the mean field 
approximation (II)}

\author{K.N.Ilinski$^{1,2}$\thanks{E-mail: kni@th.ph.bham.ac.uk}\ 
and\ A.S.Stepanenko$^{1,3}$\thanks{E-mail: ass@th.ph.bham.ac.uk}
\\
{\small\it $^{1}$ School of Physics and Space Research, University of
Birmingham,}
\\
{\small\it Birmingham B15 2TT, United Kingdom.} 
\\
{\small\it $^{2}$ Institute of Spectroscopy, Russian Academy of Sciences,} 
\\
{\small\it Troitsk, Moscow region, 142092, Russian Federation.}
\\
{\small\it $^{3}$ Theoretical Department, St-Petersburg Nuclear Physics Institute,}
\\
{\small\it Gatchina, St-Petersburg, 188350, Russian Federation.}
}
\date{ }

\begin{document}

\maketitle
\thispagestyle{empty}
\vskip -9cm
\vskip 9cm

\begin{abstract}
Self-consistent hydrodynamic one-loop quantum corrections to the Gross-Pi\-ta\-ev\-skii equation due to the interaction of the condensate with collective excitations
are calculated. 
It is done by making use a formalism of effective action and $\zeta$-function
regularization for a contribution from Bogoliubov particles in hydrodynamic approximation.
It turns out to be  possible to reduce the problem to the investigation of a determinant
of  Laplace operator on curved space where a metric is defined by density and 
velocity of the condensate. Standard methods of quantum gravity let us get the
leading logarithmic contribution of the determinant and corresponding quantum corrections.
They describe an additional quantum pressure in the condensate, local 
heating-cooling and evaporation-condensation effects of the Bose-condensed fraction. 
The effects of these corrections are studied for the correction from excited 
states in equilibrium situation. 
Response functions and form factors  are discussed in the same approximation.
\end{abstract}

\section{Introduction}
The recent success of the experimental observation of Bose-Einstein
condensation for systems of spin polarized magnetically trapped
atomic gases at ultra-low temperatures \cite{Exp1,Exp2,Exp3} and futher
investigation of their collective properties \cite{Exp4}
are stimulating
development of the theory of a nonuniform Bose condensate and its 
collective excitations.
Moreover, new exciting problems such as a  description of
the evolution of Bose condensate from relaxed trap, dynamics of 
a collapse of the condensate for Li$^7$ atoms, heating-cooling 
phenomena, various coherence effects for the 
condensate and so on are being posed   both theoretically and experimentally.

A lot of theoretical efforts were spent to solve these problems.
Evolution of the condensate from relaxed trap was considered in 
a number of papers both for symmetric and asymmetric cases \cite{HC,RHB,EDC}.
All of these calculations were performed in Hartree approximation, i.e.
based on Nonlinear Schrodinger equation or Ginzburg-Gross-Pitaevskii
equation (GPE) \cite{GP}. It is obvious that the GPE describes strongly interacting Bose condensed gas well since it is possible to ignore
non-condensed fraction. However next three points should be 
considered:
\begin{enumerate}
\item 
At least for experiments with Rb$^{87}$ number of atoms is not
such that strong interaction approximation (or Thomas-Fermi)
approximation is sufficiently accurate to produce a robust ground state profile and realistic values of experimental observables.
\item  
When the condensate expands after the release from a trap the density inevitably
decreases and quantum corrections due to the interaction with particles in
excited states  should be considered even if the original density was high
enough to be described by GPE.
\item
Effects like heating or cooling of the system and the collapse of atoms with attraction
such as Li$^7$ \cite{collapse,Stoof1,Stoof2} cannot be  described even in principle by GPE.
\end{enumerate}
Therefore excited states and their influence on the condensate require consideration. 

Firstly, excitations of a trapped system, using the technique of Bogoliubov transformation,
have been considered by Fetter \cite{F1} (see also \cite{F2}). 
 The spectrum and wave functions of excited states were derived 
by Stringari \cite{Stringari}
in the hydrodynamic approximation where the low energy
excited states are sound waves in inhomogeneous condensate media.
An analogous spectrum was obtained by
Burnett and coauthors \cite{Burnett1} who investigated response of 
the condensate to a time-dependent perturbation. Using Stringari's approach,
Wu and Griffin \cite{WG} proceeded by quantizing the low energy excitations in a trap and 
diagonalizing the hydrodynamic Hamiltonian in terms of  
normal modes associated with the amplitude and phase. The depletion of the condensate and observable quantities 
(such as response functions) are some series over all excited states. It means that they
cannot actually be evaluated analytically since contributions of various types
of excited states are the same order of magnitude (as can be seen from Fig.1
of Ref.\cite{WG}; see also discussion in section~5). Hence an effective 
method of summation  as well as the
generalization of the treatment to time-dependent and untrapped cases
are needed. These problems are solved in the present paper 
(see also \cite{IS1}).

In recent paper Ref.\cite{Kagan1} the evolution of a Bose condensed gas under
temporal variation of the confining potential was considered. It was found that
there exist exact scaling relations for a 2-dimensional gas trapped in a harmonic potential whose  frequency depends on time. Approximate relations 
 also exist
in  the 3-dimensional case if the interaction is very small or very large (such
that the kinetic energy can be neglected). This allows the authors to discuss
principal questions of the condensate evolution  from a relaxing trap and the depletion
of the condensate.
However this elegant work is restricted by the requirement of a  harmonic potential as a confining potential.

We want to note also work \cite{Stoof} which considered nucleation of the condensate.
It is possible to say that the present work is devoted to the consideration of opposite case where the condensate may evaporate.

In the paper we derive a generalization of Gross-Pitaevskii
equation  for the condensate fraction taking into account one loop quantum corrections caused by the interaction of the condensate with the
noncondensed fraction. However, there is other possibility to generalize
GP equation. Indeed, GP equation in mean field approximation
gives dynaimcal equation for both condensed fraction and full density
of the gas since the deplition is ignored in the approximation.
One loop correction for full density is obtained in Ref.\cite{IS1}
that together with the results of the present paper allows to calculate the deplition in one loop approximation simply as a difference of the full density and density of the condensate fraction.

The paper is organized as follows. In next section we consider 
the effective action and deduce the most significant terms. In Section 3 we introduce functional integration to calculate the effective
hydrodynamic action for the system and define one-loop quantum corrections which
are responsible for the interaction of the condensate with excited modes.
These quantum corrections are evaluated using $\zeta$-function regularization
for functional determinants and an explicit and very simple 
expression for the
one-loop quantum corrections is presented. This permits in Section 4 the derivation of the 
quantum corrections to the equations of motion for the condensate in closed and
explicit form readily used in numerical calculations. Here corrections 
to the Nonlinear Schr\"odinger equation (due to
the interaction of condensate with non-condensed fraction)
are obtained. Using interpolation between low and high density limits,
we show that the contribution from excitations are of order of the main terms
in a region around the classical turning points and can define 
the condensate profile there.
Section 5 is devoted to a consideration of Green functions and response functions.
It is shown that all Green functions in the 1-loop approximation can be 
obtained from the effective action of Section 3 by functional differentiation. We conclude the paper with a discussion of the possible applications of our results.  Two appendices contain basic facts and some technical details of $\zeta$-function regularizations of determinants of elliptic operators and methods of covariant variational differential calculus.

Let us emphasize that our analysis does not depend on the details of the
confining potential and can be used in a variety of problems. The only simplification  is usual hydrodynamic approximation.

\section{Exact dynamical equations for the condensate --- review}

In this section we discuss relevant contributions from the non-condensed fraction which have to include in the exact  equations of motion for the condensate. The estimate of the contributions is more  transparent if we follow the usual approach of functional integral for the condensate dynamics \cite{Popov}.

The Hamiltonian  of the system can be written as
\begin{equation}
H = \frac{\hbar^{2}}{2 m} \int\! {\rm d}V\ \left[
\nabla \psi^{+} \nabla \psi +
\upsilon (x,t)\psi^{+}\psi +
4\pi l \psi^{+}\psi^{+}\psi\psi \right],
\label{H}
\end{equation}
where $\upsilon (x,t)$ is an external potential. For the case of trapped atoms the potential is supposed to be harmonic and can be varied or even switched off at some moment, say $t=0$. More precisely it means that for the case of magnetically trapped atoms 
$\upsilon =\theta (-t) a_{\perp}^{-4}(t)(\rho^{2}+\lambda^2 (t) z^2)$, $\rho^2=x^2+y^2$, ($a_\perp$ is oscillator lengths) has to be put in Eqn(\ref{H}).
Another parameter in Eqn(\ref{H}) is $l$ --- the s-wave scattering length  
for atoms in the system \cite{Osc}.
The operators $\psi^{+},\psi$ are Bose creation and annihilation operators.
The corresponding vacuum-vacuum transition amplitude is given by 
the following functional integral:
\begin{equation}
Z(\mu) =
\int\! {\rm D}\psi^{+} (x,\tau ){\rm D}\psi (x,\tau )\ \exp[iS/\hbar] \ ,
\label{Z}
\end{equation}
with the action
\begin{equation}
S = \int_{-\infty}^{\infty} {\rm d}\tau \int\! {\rm d}V
\left\{ i\hbar\psi^{+}\frac{\partial \psi}{\partial \tau}
- \frac{\hbar^{2}}{2 m}\left[\nabla \psi^{+} \nabla \psi +
(\upsilon - \frac{2 m}{\hbar^{2}}\mu)\psi^{+}\psi + 
4\pi l \psi^{+}\psi^{+}\psi\psi \right] \right\} \ .
\label{S0}
\end{equation}
where  the parameter $\mu$ is the chemical potential, controlling the number 
of particles in the system.

As it is well-known the appearance of the Bose condensate is equivalent to the fact that besides ``normal"
quantities some ``anomalous" ones have nonzero values. The principle
``anomalous" quantities are one-point Green's functions 
$\langle\psi (t)\rangle$ and 
$\langle\psi^{+} (t)\rangle$ which are the amplitude of the condensate and have to be taken into
account accurately. The standard approach to deal with the condensate is to
make a shift of the fields $\psi^{+},\psi$ by some 
functions $\alpha^{+}, \alpha $ choosing from the condition of the vanishing 
of the ``anomalous" Green's functions (vanishing of the one-point function
provides analogies vanishing of other ``anomalous" Green's functions).
Formally, we consider the change of variables
$$
\psi^{+} = \alpha^{+} + \varphi ^{+} \qquad , \qquad 
\psi = \alpha  + \varphi \ ,
$$
which leads to a new form of the action (\ref{S0}):
\begin{equation}
S = S_{1} + S_{2} + S_{3} + S_{4} + S_{5} + S_{6}
\label{s1}
\end{equation}
\begin{eqnarray}
S_{1}\!\! &=&\!\! \int_{-\infty}^{\infty} {\rm d}\tau \int\! {\rm d}V
\left\{ 
\hbar\alpha ^{+}\frac{i\partial \alpha }{\partial \tau}
- \frac{\hbar^{2}}{2 m}
\left[\nabla \alpha ^{+} \cdot \nabla \alpha +
(\upsilon - \frac{2 m}{\hbar^{2}}\mu )\alpha^{+} \alpha  + 
4\pi l (\alpha ^{+} \alpha)^2  \right]
\right\} \ ,
\nonumber\\
S_{2} \!\! &=&\!\! \int_{-\infty}^{\infty} {\rm d}\tau \int\! {\rm d}V 
\left\{ 
\hbar\varphi ^{+} \frac{i\partial \varphi }{\partial \tau}
- \frac{\hbar^{2}}{2 m}
\left[\nabla \varphi ^{+} \cdot \nabla \varphi +
(\upsilon - \frac{2 m}{\hbar^{2}}\mu )\varphi ^{+}\varphi 
  \right] 
\right\} \ ,
\label{s2}\\
S_{3} \!\! &=&\!\! \int_{-\infty}^{\infty} {\rm d}\tau \int\! {\rm d}V
\left\{ 
\hbar\varphi ^{+}\frac{i\partial \alpha}{\partial \tau} - 
\frac{\hbar^{2}}{2 m}
\left[\nabla \varphi ^{+} \cdot \nabla \alpha +
(\upsilon - \frac{2 m}{\hbar^{2}}\mu )\varphi ^{+}\alpha +
8\pi l \varphi^{+} \alpha ^{+}\alpha \alpha \right]
\right\} \ ,
\nonumber\\
S_{4} \!\! &=&\!\! \int_{-\infty}^{\infty} {\rm d}\tau \int\! {\rm d}V
\left\{ 
\hbar\alpha^{+}\frac{i\partial \varphi}{\partial \tau}  
- \frac{\hbar^{2}}{2 m}
\left[\nabla  \alpha^{+} \cdot \nabla \varphi +
(\upsilon - \frac{2 m}{\hbar^{2}}\mu ) \alpha^{+}\varphi +
8\pi l  \alpha^{+} \alpha ^{+}\alpha\varphi \right]
\right\} \ ,
\nonumber\\
S_{5} \!\! &=&\!\! \int_{-\infty}^{\infty} {\rm d}\tau \int\! {\rm d}V
\left\{- 4\pi l\frac{\hbar^{2}}{2 m}   
\left[ 4 \alpha^{+} \alpha \varphi^{+}\varphi +
 \alpha \alpha \varphi^{+}\varphi^{+} 
+  \alpha^{+} \alpha^{+} \varphi\varphi 
\right]
\right\} \ ,
\label{s5}\\
S_{6} \!\! &=&\!\! \int_{-\infty}^{\infty} {\rm d}\tau \int\! {\rm d}V
\left\{- 4\pi l\frac{\hbar^{2}}{2 m}   
\left[ 
(\varphi ^{+}\varphi)^2  +
2\alpha \varphi^{+}\varphi^{+}\varphi 
+ 2\alpha^{+} \varphi^{+} \varphi\varphi 
\right]
\right\} \ .
\label{s6}
\end{eqnarray}
By the definition of the shifts $\alpha^{+}, \alpha$ ``normal"
quantities for $\varphi^{+},\varphi$ --- fields have to be zero,
i.e.
$$
\langle\varphi\rangle =
\int \! {\rm D}\varphi^{+} (x,\tau ){\rm D}\varphi (x,\tau )\ 
\varphi \exp[iS/\hbar] = 0
$$
and
$$
\langle\varphi^{+}\rangle =
\int\! {\rm D}\varphi^{+} (x,\tau ){\rm D}\varphi (x,\tau )\ 
\varphi^{+}\exp[iS/\hbar] = 0 \ .
$$
The last two equations are the main dynamical equations for the condensate
wave functions (shift functions):
\begin{equation}
\gamma + \Gamma = 0 \qquad , \qquad \bar{\gamma} + \bar{\Gamma}= 0
\label{eq1}
\end{equation}
where
\begin{equation}
\gamma = -i\hbar\frac{\partial \alpha}{\partial \tau}
- \frac{\hbar^{2}}{2 m} \nabla^{2} \alpha +
(\upsilon - \frac{2 m}{\hbar^{2}}\mu ) \alpha +
8\pi l  \alpha ^{+}\alpha \alpha 
\label{gamma}
\end{equation}
and $\Gamma$ is the sum of all diagrams with only one external outgoing line.
 Equation (\ref{eq1}), in the mean-field approximation (neglecting
$\Gamma$), becomes 
the Gross-Pitaevskii equation since the form of $\gamma$ (\ref{gamma}) is:
\begin{equation}
-i\hbar\frac{\partial \alpha}{\partial \tau}
- \frac{\hbar^{2}}{2 m} \nabla^{2} \alpha +
(\upsilon - \frac{2 m}{\hbar^{2}}\mu ) \alpha +
8\pi l  \alpha ^{+}\alpha \alpha  = 0 \ .
\label{GP}
\end{equation}
In  previous work on Bose condensates in traps  only equation (\ref{GP}) has been studied.

Let us now estimate corrections  to $\Gamma$ from the interaction of the
condensate with the non-condensed fraction. 
For the case of magnetically trapped atoms, i.e. a
harmonic potential $\upsilon= a_{\perp}^{-4}(\rho^{2}+\lambda^2 z^2)$ 
( $\lambda$ is the
anisotropy parameter) there is natural unit of length
in the system, which is $a_{\perp}$. After rescaling the coordinates
in units $a_{\perp}$, the interaction constant
$4\pi l$ is changed to $g = 4\pi l/a_{\perp}$ (in experimental conditions
$g\sim 0.05$). Starting with the mean field solution for the condensate
fraction in the trap, we can estimate effective interactions
(\ref{s5}) and (\ref{s6}). Indeed,
the mean field solution for a number of particles 
$N \sim 5000$ at the center of the trap is of order of 10.
From this it follows that at the initial stage, the main
contribution to the Gross-Pitaevskii equation is the term
$2g|\alpha|^2 \alpha \sim 100$. However the next contribution from the interaction with non-condensed excited states with vertex  
$4g |\alpha|^2 \sim 20 \gg 1$ cannot be neglected. By comparison
all other terms are quite small ($g |\alpha | \sim 0.5$, $g\sim 0.05$) and
correspond to the weakly interacting non-condensed Bose gas contributions
(for the original experimental setting with $N\sim 2000$ the numbers
are $2g|\alpha|^2 \alpha \sim 30$, $4g|\alpha|^2 \sim 10$, $g |\alpha | \sim 0.3$, $g\sim 0.05$). Hence the description of the transition stage from
the original trapped condensate to the dilute phase require taking into account interaction effects due to terms (\ref{s5}) while the interactions
(\ref{s6}) can be ignored. Moreover, cooling-heating effects, dissipation
from the condensate and so on come from the interaction terms (\ref{s5}) and
cannot be neglected in the resulting description of the dynamics of the condensate.

It is easy to see that the interaction (\ref{s5}) together with the quadratic
part (\ref{s2}) correspond to a Gaussian approximation in the integration over oscillations around mean field solutions, i.e. Bogoliubov's excitations.
It is quite hard to integrate over Bogoliubov's excitations in general but
this significantly simplifies in the hydrodynamic limit. As it was shown in
\cite{WG} it is possible to obtain  equations for hydrodynamic excitations straightforward from Bogoliubov's equations and then second quantize the excitations.
Instead we will derive the hydrodynamic description directly from the functional integral formalism, which simplifies the description.

\section{Quantum corrections to mean field equation in hydrodynamic
regime}

We are 
looking for the effective action which describes all physical quantities for the system (for easy introduction to the formalism of the effective action see, for example, Appendix A of Ref.\cite{IS1}). For example, the effective action provides all
Green's functions in the same approximation used to calculate the effective action itself. We will obtain it in a hydrodynamic
one-loop approximation. To clarify the
description hydrodynamical variables, density and velocity, should be used.
More precisely, we start with the action (\ref{S0})
\begin{equation}
\frac{S}{\hbar} = \frac{1}{\hbar}\int_{-\infty}^{\infty} 
{\rm d}\tau \int\! {\rm d}V
\left\{i\hbar \psi^{+} \frac{\partial \psi}{\partial \tau}
- \frac{\hbar^{2}}{2 m}
\left[\nabla \psi^{+} \nabla \psi +
(\upsilon - \frac{2 m}{\hbar^{2}}\mu)\psi^{+}\psi + 
4\pi l (\psi^{+}\psi)^2 \right] \right\} \ ,
\end{equation}
and rescale variables as $\tau \rightarrow \frac{2m}{\hbar}\tau$, 
$\mu \rightarrow \frac{\hbar^2}{2m} \mu$, 
$l \rightarrow \frac{l}{4\pi}$
to get the following action form:
\begin{equation}
\frac{S}{\hbar} = \int_{-\infty}^{\infty} {\rm d}\tau \int\! {\rm d}V
\left\{ i\psi^{+} \frac{\partial \psi}{\partial \tau}
- \left[\nabla \psi^{+} \nabla \psi +
(\upsilon  - \mu)\psi^{+}\psi + 
 l (\psi^{+}\psi)^2 \right] \right\} \ .
\label{S1}
\end{equation}
(We will measure the action in terms of $\hbar$ everywhere below).
Now we change field variables to the hydrodynamic ones:
\begin{equation}
\psi (x,\tau) = \sqrt{\rho(x,\tau)} e^{-i\varphi(x,\tau)}
\qquad , \qquad 
\psi^+ (x,\tau) = \sqrt{\rho(x,\tau)} e^{i\varphi(x,\tau)} \ .
\label{psi}
\end{equation}
Then the action (\ref{S1}) takes the form (up to a complete derivative term):
\begin{equation}
S = \int_{-\infty}^{\infty} {\rm d}\tau \int\! {\rm d}V
\left\{ 
\frac{\partial \varphi}{\partial \tau}\rho -
(\nabla \sqrt{\rho})^2 - \rho (\nabla \varphi)^2 -
(\upsilon - \mu)\rho -
 l \rho^2 \right\} \ .
\label{S2}
\end{equation}
It is not difficult to see that the classical equations of motion for the
action lead to the equations:
\begin{equation}
\frac{\partial \varphi}{\partial \tau} -
 (\nabla \varphi)^2
-v  + \mu -
 2 l \rho + \frac{1}{\sqrt{\rho}}\nabla^2 \sqrt{\rho}  = 0 \ ,
\label{eqn1}
\end{equation}
\begin{equation}
- \frac{\partial \rho}{\partial \tau} +
2 \nabla (\nabla\varphi  \cdot\rho)
  = 0 \ ,
\label{eq2}
\end{equation}
which are equivalent to the Gross-Pitaevskii equation \cite{Stringari}
after the introduction of the velocity variable ${\bf c}=-2\nabla \varphi$
instead of $\varphi$. In velocity-density variables these equations look
as hydrodynamic equations for an irrotational compressible fluid:
\begin{equation}
\frac{\partial {\bf c}}{\partial \tau} + \nabla \left(
\frac{{\bf c}^2}{2}
+2v  - 2\mu +
4 l \rho - \frac{2}{\sqrt{\rho}}\nabla^2 \sqrt{\rho}\right)  = 0 \ ,
\label{eq3}
\end{equation}
\begin{equation}
\frac{\partial \rho}{\partial \tau} +
 \nabla ({\bf c}  \cdot\rho)
  = 0 \ .
\label{eq4}
\end{equation}

Now let us shift our variables in (\ref{S1}) by the zero-order mean field solution
to find one loop quantum correction for the effective action:
$$
\psi \rightarrow \psi + \chi \qquad , \qquad 
\psi^{+} \rightarrow \psi^{+} + \chi^{+}
$$
and we will keep up only terms including the square of $\chi$ and $\chi^{+}$ 
terms. Then the action 
(\ref{S1}) transforms to $S(\psi,\psi^{+}) + S_1(\psi,\psi^{+},\chi,\chi^{+})$
where the action $S_1$ has the form:
\begin{eqnarray}
S_1(\psi,\psi^{+},\chi,\chi^{+}) &=& 
\int_{-\infty}^{\infty} {\rm d}\tau \int\! {\rm d}V
\Biggl\{ i\chi^{+} \frac{\partial \chi}{\partial \tau}
- \Bigl[\nabla \chi^{+} \nabla \chi +
\{\dot{\varphi} - (\nabla\varphi)^2 + 2l\rho \}\chi^{+}\chi
\nonumber\\
 &&\quad - l {\rm e}^{2i\varphi}\rho \chi^{+} \chi^{+}
 - l {\rm e}^{-2i\varphi}\rho \chi \chi 
\Bigl] \Biggl\} \ .
\label{S3}
\end{eqnarray}
where hydrodynamic representation (\ref{psi}) were used and potential $\upsilon$
was expressed through mean field equation of motion neglecting kinetic term.
Hamiltonian corresponding to $S_1$ is
\begin{equation}
H = \int\! {\rm d}V
\left[\nabla \hat\chi^{+} \nabla \hat\chi +
\{ \dot{\varphi} - (\nabla\varphi)^2 + 2l\rho \} \hat\chi^{+}\hat\chi 
 - l {\rm e}^{2i\varphi}\rho \hat\chi^{+} \hat\chi^{+}
 - l {\rm e}^{-2i\varphi}\rho \hat\chi \hat\chi 
\right] \ .
\label{Ham}
\end{equation}
Now we fulfill the following canonical transformation:
$$
\hat\chi = 
\sqrt{\rho}{\rm e}^{i\varphi}\left[\frac{\hat\sigma}{2\rho} + i\hat\alpha\right]
\qquad , \qquad
\hat\chi^{+} = 
\sqrt{\rho}{\rm e}^{-i\varphi}\left[\frac{\hat\sigma}{2\rho} - i\hat\alpha\right]
$$
The hamiltonian takes the form
\begin{eqnarray}
H &=& \int\! {\rm d}V
\Biggl[
   \rho (\nabla\hat\alpha)^2
 + \rho\dot\varphi \hat\alpha^2
 + \{ l + \frac{\displaystyle\dot\varphi}{\displaystyle 4\rho} \} \hat\sigma^2
 + \hat\sigma (\nabla\varphi\nabla\hat\alpha)
 - \hat\alpha (\nabla\varphi\nabla\hat\sigma)
\nonumber\\
&&
 + \frac{1}{2}\Delta\varphi\hat\sigma\hat\alpha 
 - \frac{1}{2}\Delta\varphi\hat\alpha\hat\sigma 
 + \frac{\displaystyle (\nabla\rho)^2}{\displaystyle 4\rho}\hat\alpha^2
 + \frac{1}{\displaystyle 4\rho}(\nabla\hat\sigma)^2
 + \frac{\displaystyle \nabla\rho\nabla\varphi}{\displaystyle \rho}
	\hat\alpha\hat\sigma
\nonumber\\
&&
 - \frac{1}{\displaystyle 4\rho^2}\hat\sigma\nabla\rho\nabla\hat\sigma
 + \frac{\displaystyle (\nabla\rho)^2}{\displaystyle 16\rho^3}\hat\sigma^2
 + \hat\alpha\nabla\rho\nabla\hat\alpha
\Biggr] \ .
\label{Ham1}
\end{eqnarray}
Last eight terms  vanish in the hydrodynamical limit since
$\rho$ is a large variable ($\nabla\varphi\sim\sqrt{l\rho}$ 
and $\dot\varphi\sim l\rho$). So we have
\begin{equation}
H = \int\! {\rm d}V
\Biggl[
   \rho (\nabla\hat\alpha)^2
 + \rho\dot\varphi \hat\alpha^2
 + \{ l + \frac{\displaystyle\dot\varphi}{\displaystyle 4\rho} \} \hat\sigma^2
 + \hat\sigma (\nabla\varphi\nabla\hat\alpha)
 - \hat\alpha (\nabla\varphi\nabla\hat\sigma)
\Biggr] \ .
\label{Ham2}
\end{equation}
The corresponding action (in feynmann functional integral) is
\begin{equation}
S_1 = 
\int_{-\infty}^{\infty} {\rm d}\tau \int\! {\rm d}V
\Biggl\{ \frac{\partial \alpha}{\partial \tau}\sigma
 - \Bigl[
   \rho (\nabla\alpha)^2
 + \rho\dot\varphi \alpha^2
 + \{ l + \frac{\displaystyle\dot\varphi}{\displaystyle 4\rho} \} \sigma^2
 + 2\sigma (\nabla\varphi\nabla\alpha)
\Bigr]
\Biggr\} \ .
\label{S4}
\end{equation}
Integrating in functional integral over the  field $\sigma$ we get  the action for 
$\alpha$ field only:
\begin{equation}
S_1 = \int_{-\infty}^{\infty} {\rm d}\tau \int\! {\rm d}V
\left\{ -\rho (\nabla \alpha)^2 +
f\Bigl(\frac{\partial\alpha}{\partial \tau} 
 - 2 \nabla \varphi \nabla \alpha 
\Bigr)^2
 - \rho\dot\varphi\alpha^2 
\right\} \ .
\label{S5}
\end{equation}
where
$$
f = \frac{1}{4l + \dot\varphi/\rho}\ .
$$
The corresponding determinant does not contribute.

Now we will use the large  parameter $\rho$.
Let us introduce the following large number 
$\rho_0 \equiv {\rm max}\, \{\rho \}$ such that
$\tilde{\rho} = \frac{\rho}{\rho_0} \sim 1$ and new space-time variables 
$t \equiv 4l\rho_{0}\tau $, $y_i\equiv \sqrt{ 4l\rho_{0}} x_i$. Hence
\begin{eqnarray}
S_1\!\!& =&\!\! 
\frac{1}{\sqrt{64\rho_0 l^3}} 
\int_{-\infty}^{\infty} {\rm d}t \int\! {\rm d}\tilde V\
\left\{ -\tilde{\rho} (\tilde\nabla \alpha)^2 +
\tilde f\Bigl(
\frac{\partial\alpha}{\partial t}
 - \tilde{\bf v} \tilde\nabla \alpha\Bigr)^2
 - \tilde\rho\dot\varphi\alpha^2 
\right\} 
\nonumber\\
\!\!& \equiv&\!\!
 - \frac{1}{\sqrt{64\rho_0 l^3}}
\int_{-\infty}^{\infty} {\rm d}t \int\! {\rm d}\tilde V\
\left[
A^{\mu \nu} \partial_{\mu}\alpha \partial_{\nu}\alpha
 + \tilde\rho\dot\varphi\alpha^2
\right]\ .
\label{S6}
\end{eqnarray}
where
$$
\tilde{\bf v} = 2\tilde\nabla \varphi\ , \quad
\tilde f = 4lf = \frac{1}{1 + \dot\varphi/\tilde\rho}
$$
and the matrix $A$ has a form
$$
A = 
\left(  
\begin{array}{cccc}
-\tilde f & \tilde f\tilde{v}_1  & \tilde f\tilde{v}_2  &\tilde f \tilde{v}_3 \\
\tilde f\tilde{v}_1 & \tilde{\rho}-\tilde f\tilde{v}_1^2
	 & -\tilde f\tilde{v}_1\tilde{v}_2 &  -\tilde f\tilde{v}_1\tilde{v}_3 \\
\tilde f\tilde{v}_2 & -\tilde f\tilde{v}_1\tilde{v}_2 
	 & \tilde{\rho}-\tilde f\tilde{v}_2^2 &  -\tilde f\tilde{v}_2\tilde{v}_3 \\
\tilde f\tilde{v}_3 & -\tilde f\tilde{v}_1\tilde{v}_3 
 	 & -\tilde f\tilde{v}_2\tilde{v}_3 & \tilde{\rho}-\tilde f\tilde{v}_3^2  \\
\end{array} 
\right)
$$
with its determinant ${\rm det}\, A = -\tilde{\rho}^{3}\tilde f$.

Our next step is to cast the action in the covariant form. To do this we introduce
auxiliary metric $\tilde{g}_{\mu\nu}$ such that 
$$
A^{\mu \nu} = \frac{\tilde{g}^{\mu\nu}}{\sqrt{-{\rm det}\, 
(\| \tilde{g}^{\mu\nu} \|)}} \ .
$$
One can easy to find the covariant metric from the equation above:
$$
\tilde{g}^{\mu \nu} = \frac{A^{\mu\nu}}{\sqrt{-{\rm det}\, (\| A^{\mu\nu} \|)}}
$$
that leads to the form:
$$
\|\tilde{g}^{\mu\nu}\|= \tilde{f}^{-1/2}\tilde{\rho}^{-3/2}
\left(
\begin{array}{cccc}
-\tilde f & \tilde f\tilde{v}_1  & \tilde f\tilde{v}_2  &\tilde f \tilde{v}_3 \\
\tilde f\tilde{v}_1 & \tilde{\rho}-\tilde f\tilde{v}_1^2
	 & -\tilde f\tilde{v}_1\tilde{v}_2 &  -\tilde f\tilde{v}_1\tilde{v}_3 \\
\tilde f\tilde{v}_2 & -\tilde f\tilde{v}_1\tilde{v}_2 
	 & \tilde{\rho}-\tilde f\tilde{v}_2^2 &  -\tilde f\tilde{v}_2\tilde{v}_3 \\
\tilde f\tilde{v}_3 & -\tilde f\tilde{v}_1\tilde{v}_3 
 	 & -\tilde f\tilde{v}_2\tilde{v}_3 & \tilde{\rho}-\tilde f\tilde{v}_3^2  \\
\end{array}  
\right) \ .
$$
Finally we get the metric $\|\tilde{g}_{\mu\nu}\|$:
$$
\|\tilde{g}_{\mu\nu}\|= \tilde{\rho}^{1/2}\tilde{f}^{1/2}
\left(  
\begin{array}{cccc}
-\tilde{\rho}/\tilde f + \tilde{v}^2 & \tilde{v}_1 & \tilde{v}_2 & \tilde{v}_3 \\
\tilde{v}_1 & 1 & 0  &  0 \\
\tilde{v}_2 & 0 &  1 &  0  \\
\tilde{v}_3 & 0 & 0 &  1  \\
\end{array} \right) \ .
$$
with the determinant 
$\tilde{g}\equiv {\rm det}\,(\|\tilde{g}_{\mu\nu}\|) = -\tilde{\rho}^{3}\tilde f$.

In this metric the action takes a covariant form
\begin{eqnarray}
S_1 &=&
 -\frac{1}{\sqrt{64\rho_0 l^3}}
\int\! {\rm d}y\ \sqrt{-\tilde{g}}
 \left[\alpha  \tilde{\mathop\Box^{ }} \alpha + \tilde E \alpha^2\right] 
\nonumber\\
 &=& \frac{1}{\sqrt{64\rho_0 l^3}}
\int\! {\rm d}y\ 
\sqrt{-\tilde{g}} 
 \alpha \left[
\frac{1}{\sqrt{-\tilde{g}}} \partial_{\mu} \tilde{g}^{\mu \nu}\sqrt{-\tilde{g}} 
\partial_{\nu} + \tilde E \right]\alpha \ ,
\label{S8}
\end{eqnarray}
where
$$
\tilde E = \frac{\tilde \rho\dot\varphi}{\sqrt{-\tilde g}}
 = \frac{\dot\varphi}{\sqrt{\tilde\rho\tilde f}}\ .
$$
This is the central equation of the paper.

Now the effective action is 
$\Gamma _{1} = - \frac{1}{2}{\rm tr}\, \ln [(64\rho_0 l^3)^{-1/2} 
(\tilde{\displaystyle\mathop\Box^{ }}+\tilde E)]$
with Laplace operator 
$\tilde{\displaystyle\mathop\Box^{ }} 
= -\frac{1}{\sqrt{-\tilde{g}}} \partial_{\mu} 
\tilde{g}^{\mu \nu}  \sqrt{-\tilde{g}} \partial_{\nu} $.
Let us again make use the existence of large parameter in the system.
Indeed, the multiplier $(64\rho_0 l^3)^{-1/2}$ gives a possibility to
express main contributions to the determinant such as (see Appendix A):
$$
{\rm tr}\, \ln [(64\rho_0 l^3)^{-1/2} 
(\tilde{\displaystyle\mathop\Box^{ }}+\tilde E)]
 = {\rm tr}\, \ln [ 
\tilde{\displaystyle\mathop\Box^{ }}+\tilde E ]
 - \frac{1}{2} {\rm tr}\, \ln (64\rho_0 l^3) 
\Bigl(\Phi_0 (\tilde{\displaystyle\mathop\Box^{ }}+\tilde E )
 - L(\tilde{\displaystyle\mathop\Box^{ }}+\tilde E) \Bigr)
$$
$$
\sim 
 - \frac{1}{2} {\rm tr}\, \ln (64\rho_0 l^3) 
\Bigl(\Phi_0 (\tilde{\displaystyle\mathop\Box^{ }}+\tilde E )
 - L(\tilde{\displaystyle\mathop\Box^{ }}+\tilde E) \Bigr)
$$
since in our regularization 
${\rm tr}\, \ln [ \tilde{\displaystyle\mathop\Box^{ }}+\tilde E]$ is order of
the zeroth Seeley 
coefficient $\Phi_0(\tilde{\displaystyle\mathop\Box^{ }}+\tilde E)\equiv
 \int\!  {\rm d}y{\rm d}\tau\ \sqrt{-\tilde g} \Psi_0$.
We will consider only cases where the number of zero-modes does not change.
So in what follows we will drop the term with 
$L(\tilde{\displaystyle\mathop\Box^{ }}+\tilde E)$.
Returning to the initial variables all curvature tensors  and the metric
are written in the ``physical" variables (i.e. without tildes).
Moreover $\ln (64\rho_0 l^3) \gg \ln (\tilde{\rho})$ so that we can substitute
$\ln (64\rho l^3)$ instead of $\ln (64\rho_0 l^3)$. Summarizing, 
we obtain an expression for the first quantum correction to the
effective action:
\begin{equation}
\Gamma _{1} = \frac{1}{4} \int\!  {\rm d}x{\rm d}\tau\ \sqrt{-g}
\ln (64\rho l^3) \Psi_0(\Box+E) \ ,
\label{Gamma1}
\end{equation}
with the metric 
$$
\|g_{\mu\nu}\|= \rho^{1/2}f^{1/2}
\left(  
\begin{array}{cccc}
-\rho/f + v^2 &  v_1  &  v_2  &  v_3  \\
v_1 & 1 & 0  &  0 \\
v_2 & 0 &  1 &  0  \\
v_3 & 0 & 0 &  1  \\
\end{array} \right) \ .
$$
and 
$$
v_i\equiv 2\partial_i\varphi\ ,\quad 
f = \frac{1}{4l + \dot\varphi/\rho}\ ,\quad
E = \frac{\dot\varphi}{\sqrt{\rho f}}\ .
$$

\section{Quantum corrections to equations of motion}

In this section we derive from (\ref{Gamma1}) required quantum corrections to equations of motion (see, for example, \cite{IS1}).
To make the consideration self-consistent  we should not differentiate 
$ \ln (64\rho l^3)$ because in our approximation it behaves like a constant.
In that case, the action we have to vary is covariant again.
We find:
$$
\frac{\delta \Gamma_{1}}{\delta \rho} = 
\frac{1}{4} \ln (64\rho l^3)
\Biggl[
\frac{\partial g_{\mu\nu}}{\partial \rho}
\frac{\delta \Phi_{0}}{\delta g_{\mu\nu}}
 + \frac{\partial E}{\partial \rho} 
\frac{\delta \Phi_{0}}{\delta E}
\Biggr]\ ,
$$
$$
\frac{\delta \Gamma_{1}}{\delta \varphi} = 
\frac{1}{4} \ln (64\rho l^3)
\int\!  {\rm d}^4 y\ 
\Biggl[
\frac{\delta g_{\mu\nu}(x)}{\delta \varphi(y)}
\frac{\delta \Phi_{0}}{\delta g_{\mu\nu}(x)}
 + \frac{\delta E(x)}{\delta \varphi(y)} 
\frac{\delta \Phi_{0}}{\delta E(x)}
\Biggr] \ .
$$
Let us remember that
\begin{eqnarray*}
\Phi_{0} (\Box + E) \!\!&=\!\!&
\frac{1}{(4\pi)^2} \int \!{\rm d}x\ \sqrt{-g}\
\Biggl[ 
 - \frac{1}{30}\nabla^2 R
 + \frac{1}{72}R^2
 - \frac{1}{180}R_{\mu\nu} R^{\mu\nu}
\\
\!\!&\!\!&
 + \frac{1}{180}R_{\mu\nu\sigma\rho} R^{\mu\nu\sigma\rho} 
 + \frac{1}{6}RE
 + \frac{1}{2} E^2 
 - \frac{1}{6}\nabla^2 E
\Biggr] 
\end{eqnarray*}
We note that two of the terms above are irrelevant:
$$
\int \!{\rm d}x\ \sqrt{-g} \nabla^2 F = 
\int \!{\rm d}x\ \sqrt{-g} \frac{1}{\sqrt{-g}}
\partial_{\mu} \sqrt{-g} g^{\mu\nu} \partial_{\nu} F
 = \int \!{\rm d}x\  \partial_{\mu} (\sqrt{-g}\partial^{\mu} F)
$$
since this is just a complete derivative. Hence for our purposes it is sufficient to consider only 
$$
\Phi_{0} (\Box + E) =
\frac{1}{(4\pi)^2} \int \!{\rm d}x\ \sqrt{-g}\
\Biggl[ 
   \frac{1}{72}R^2
 - \frac{1}{180}R_{\mu\nu} R^{\mu\nu}
 + \frac{1}{180}R_{\mu\nu\sigma\rho} R^{\mu\nu\sigma\rho} 
 + \frac{1}{6}RE
 + \frac{1}{2} E^2 
\Biggr] 
$$

Long but quite straightforward calculations lead us to the following result for the
functional derivative (see Appendix B for details of the calculation):
\begin{eqnarray*}
\frac{\delta }{\delta g^{\mu\nu}} \Phi_{0} 
\!\!& =\!\! &
 \frac{1}{(4\pi)^2}
\sqrt{-g} 
\Biggl\{
 - \frac{1}{2} g_{\mu\nu} 
\Bigl[ 
   \frac{1}{72}R^2
 - \frac{1}{180} R_{\sigma\rho}R^{\sigma\rho} 
 + \frac{1}{180}R_{\sigma\rho\alpha\beta}R^{\sigma\rho\alpha\beta}
+ \frac{1}{6} RE + \frac{1}{2} E^2
\Bigr] 
\\
\!\!& +\!\! & 
\frac{1}{6} R_{\mu\nu}E
  + \frac{1}{36} R_{\mu\nu}R
 - \frac{1}{90}R_{\mu\sigma}R_{\nu}^{\,\cdot\,\sigma} 
 + \frac{1}{90}R_{\mu\sigma\rho\alpha}R_{\nu}^{\,\cdot\,\sigma\rho\alpha}
 - \frac{1}{36}g_{\mu\nu} \Box R
\\
\!\!& +\!\! & 
\frac{1}{72} \{\nabla _{\mu},\nabla_{\nu}\} R  
 - \frac{1}{6}g_{\mu\nu} \Box E 
 + \frac{1}{12} \{\nabla _{\mu},\nabla_{\nu}\} E
\\
\!\!& +\!\! & 
\frac{1}{180} 
\Bigl[ 
 - 2\nabla^{\sigma}\nabla_{\nu} R_{\mu\sigma}
 + \Box R_{\mu\nu}
 + g_{\mu\nu}\nabla^{\sigma}\nabla^{\rho} R_{\sigma\rho} 
\Bigr]
 + \frac{1}{45} 
   \nabla^{\rho}\nabla^{\sigma}R_{\mu\sigma\rho\nu} 
\Biggr\}
\end{eqnarray*}
taking into account
$$
\frac{\delta }{\delta g_{\mu\nu}} \Phi_{0} (\Box + E) =
- g^{\mu\sigma}g^{\nu\rho}\frac{\delta }{\delta g^{\sigma\rho}} 
\Phi_{0} (\Box + E)  \ .
$$
This means that the equations of motions with 1-loop quantum correction 
take the form:
\begin{equation}
\frac{\partial \varphi}{\partial \tau}
 - (\nabla \varphi)^2
 - v  + \mu
 - 2 l \rho
 + \frac{1}{4} \ln (64\rho l^3)
\Biggl[
\frac{\partial g_{\mu\nu}}{\partial \rho}
\frac{\delta \Phi_{0}}{\delta g_{\mu\nu}}
 + \frac{\partial E}{\partial \rho} 
\frac{\delta \Phi_{0}}{\delta E}
\Biggr]
 = 0 \ ,
\label{eqncorr1}
\end{equation}
\begin{equation}
 - \frac{\partial \rho}{\partial \tau}
 + 2 \nabla (\nabla\varphi  \cdot\rho)
 + \frac{1}{4} \ln (64\rho l^3)
\Biggl[
\int\!  {\rm d}^4 y\ 
\frac{\delta g_{\mu\nu}(x)}{\delta \varphi(y)}
\frac{\delta \Phi_{0}}{\delta g_{\mu\nu}(x)}
 - \frac{\partial}{\partial\tau}
\left(
\frac{\partial E}{\partial \dot\varphi} 
\frac{\delta \Phi_{0}}{\delta E}
\right)
\Biggr] 
= 0 \ ,
\label{eqncorr2}
\end{equation}
These two equations (\ref{eqncorr1}) and (\ref{eqncorr2}) replace the mean field
hydrodynamic equations (\ref{eqn1},\ref{eq2}) for the density and velocity of the
condensate and are the main result of the paper. They contain contributions
of the excited states to the dynamics of the condensed fraction and provide  a
description of effects such as depletion and heating-cooling. For example, we
 note that the additional quantum pressure is due to dependence of the
Bogoliubov particles determinant on the density of the condensate while
an evaporation of the condensate (see continuity equation (\ref{eqncorr2})) comes
from phase-dependence of the determinant. In particular, for the stationary
situation when the equilibrium condensate velocity is equal to zero there is no
evaporation of the condensate and the continuity equation holds for the 
fraction. This case was considered in Ref.\cite{IS1}. However we think that 
equations (\ref{eqncorr1}) and (\ref{eqncorr2}) are of the main
interest in non-equilibrium problems such as evolution of the condensate under
changing of  trap shapes,  collapse of condensate cloud and corresponding heating
and evaporation of the condensate and so on. We hope to consider these problems in forthcoming papers.

\section{Green functions, response functions and the like}

Equations of motion and quantum corrections for them are the central questions of previous consideration.  However in many applications related to experiment
it is very important to analyze  various response functions and form-factors.
They can be expressed in terms of  some combinations of Green functions of the theory.
That is why in this section we shortly consider a calculation of Green functions in the effective action formalism. 

As it is shown in Ref \cite{IS1} the effective action (in any self-consistent approximation) allows the evaluation of {\it all} Green functions in the same
approximation by just taking of variational derivatives. Since the effective action
in 1-loop approximation was obtained the problem of Green function calculation
is the problem of functional differentiation only \cite{note3}.

We now formalize all said above and give formulas to evaluate Green functions
in effective action formalism.
Let $W(J)$ being generating functional of connected Green functions. Then 
quantities
\begin{equation}
        W_n(x_1,\dots,x_n) = \frac{\delta}{\delta J(x_1)}\dots
        \frac{\delta}{\delta J(x_n)}\,W(J)
\label{G1}
\end{equation}
are connected Green functions in the external field $J$. The effective action 
is defined then by the Legendre transformation of $W$:
\begin{equation}
        \Gamma(\alpha) = W(J(\alpha)) - \alpha J(\alpha)\ ,\qquad
        \alpha = \frac{\delta W(J)}{\delta J}
\label{G2}
\end{equation}
where the function
$\alpha(x) = \langle\hat\varphi(x)\rangle = W_1(x;J)$ is the first 
connected Green function (\ref{G1}). It is easy
to see that the functions $\alpha$ and $A$ are related by the 
second relation (\ref{G2}). Then the quantities
\begin{equation}
        \Gamma_n(x_1,\dots,x_n) = \frac{\delta}{\delta \alpha(x_1)}\dots
        \frac{\delta}{\delta \alpha(x_n)}\,\Gamma(\alpha)
\label{G3}
\end{equation}
define 1PI (one-particle-irreducible) Green functions for the theory with
mean field $\alpha(x) = \langle\hat\varphi(x)\rangle$. Knowledge of 1PI Green function is equivalent to knowledge of any (whole or connected) Green functions for the
corresponding system. For the first 1PI 
Green function we have from (\ref{G2})
\begin{equation}
        \Gamma_1(x) = - J(x) \ .
\label{G4}
\end{equation}
All connected Green functions can be expressed in terms of 1PI ones. 
Indeed, differentiating (\ref{G4}) $J$ we obtain
\begin{equation}
        W_2\Gamma_2 = - 1\ ,\qquad W_2 = - \Gamma_2^{-1}
\label{G5}
\end{equation}
or, in expanded form,
\begin{equation}
        \int\!{\rm d}z\ W_2(x,z)\Gamma_2(z,y) = - \delta(x-y) \ .
\label{IP}
\end{equation}
Differentiating now the second relation in (\ref{G5}) on $J$ and using the 
following rule
$$
        \frac{\delta}{\delta J} = - \Gamma_2^{-1} \frac{\delta}{\delta \alpha}
$$
we can derive expressions for all the higher connected Green functions through
the 1PI ones. For example, for the third connected function we have
$$
        W_3 = - \left[\Gamma_2^{-1}\right]^{3}\Gamma_3
$$
or
\begin{equation}
        W_3(x_1,x_2,x_3) = - \int\!{\rm d}y_1{\rm d}y_2{\rm d}y_3\ 
        \Gamma_2^{-1}(x_1,y_1)\Gamma_2^{-1}(x_2,y_2)
        \Gamma_2^{-1}(x_3,y_3)\Gamma_3(y_1,y_2,y_3)
\label{IP2}
\end{equation}
and so on. It means that since we know 1PI functions we have to solve
the differential equation (\ref{IP}) for the two-point correlation function
and then find all other connected Green functions by integration of 1PI functions with
two-point correlators as in Eqn.(\ref{IP2}).

Putting $J=0$ in (\ref{G4}) we get equations of motion for $\psi$ and 
$\psi ^+$ (\ref{eqncorr1},\ref{eqncorr2}). To obtain, for example, the Green function 
$\langle(\psi^+(x)-\langle\psi^+ (x)\rangle )(\psi(y) - \langle \psi (y)\rangle ) \rangle$ we twice differentiate the effective
action on $\psi , \psi^+$, substitute solution of equations of motion and finally  invert the result in the sense of the kernel of a integral operator.

As an example of usage of this technique we calculate two-point Green function in the mean field approximation (tree or 0-loop approximation). 
In this approximation the effective action has the form:
\begin{equation}
\Gamma = \int_{-\infty}^{\infty} {\rm d}\tau \int\! {\rm d}V
\left\{ 
-i\frac{\partial \psi^+}{\partial \tau}\psi -
\nabla \psi^{+} \nabla \psi -
(\upsilon - \mu)\psi^{+}\psi -
 l (\psi^{+}\psi)^2 \right\} \ .
\label{ac0}
\end{equation}
such that, in the equilibrium hydrodynamic picture, the matrix of second variational derivatives
of the effective action can be written as
$$
\delta^2 \Gamma = 
\left(  
\begin{array}{cc}
\displaystyle\frac{\delta^2 \Gamma}{ \delta \psi^+ (x) \delta \psi^+ (y)}  &  
\displaystyle\frac{\delta^2 \Gamma}{ \delta \psi^+ (x) \delta \psi (y)}
 \\[3mm]
\displaystyle\frac{\delta^2 \Gamma}{  \delta \psi (x)  \delta \psi^+ (y)}  &  
\displaystyle\frac{\delta^2 \Gamma}{ \delta \psi (x) \delta \psi (y)} \\
\end{array} 
\right)
= 
$$
$$  =
\left(  
\begin{array}{cc}
-2 l \psi^2 &  
\displaystyle -i\frac{\partial }{\partial \tau} -
\nabla^2 -
(\upsilon - \mu) -
 2 l (\psi^{+}\psi)\\
\displaystyle -i\frac{\partial }{\partial \tau} -
\nabla^2 -
(\upsilon - \mu) -
 2 l (\psi^{+}\psi) &
 -2 l (\psi^+)^2 \\
\end{array} 
\right)  \  .
$$
where $\psi^{(+)}\equiv\langle\hat\psi^{(+)}\rangle$. The
matrix of two-point connected correlation functions of fluctuations
$\hat{\tilde\psi} (x) = \hat\psi (x) - \langle \hat\psi (x) \rangle$ 
and 
$\hat{\tilde\psi^+} (x)= \hat\psi^+ (x) - \langle \hat\psi^+ (x) \rangle$
$$
G = 
\left(  
\begin{array}{cc}
\langle \hat{\tilde{\psi^+} } (x) \hat{\tilde{\psi^+} } (y) \rangle  &  
\langle \hat{\tilde{\psi^+} } (x) \hat{\tilde{\psi} } (y) \rangle \\[1.5mm]
\langle \hat{\tilde{\psi} } (x) \hat{\tilde{\psi^+} } (y) \rangle  &  
\langle \hat{\tilde{\psi} } (x) \hat{\tilde{\psi} } (y) \rangle \\
\end{array} 
\right)
$$
is defined as a solution of the equation
$$
\delta^2\Gamma  \cdot  G  = - {\rm I} \ .
$$
Solving the equation we get the following expressions for the correlators.
One can check that the Green functions calculated in this manner coincide with 
the Green functions $\langle \hat{\tilde\rho } (x) \hat{\tilde\rho } (y) \rangle$, 
$\langle \hat{\tilde\rho } (x) \hat{\tilde\varphi } (y) \rangle, 
\langle \hat{\tilde\varphi } (x)  \hat{\tilde\varphi } (y) \rangle$
analyzed in spectral representation by Wu and Griffin \cite{WG}. 
To include the first quantum correction 
(1-loop quantum correction)  term (\ref{Gamma1}) should be added to action (\ref{ac0}) \cite{note2}.

In summary, various response functions, form-factors and Green functions
are evaluated from effective action in any given approximation taking
variational derivatives and integrating them with the two-point correlation function. This can be done in general  using formulas of this section.

\section{Conclusion}

In the paper a self-consistent approach to the calculation of
1-loop quantum corrections to the Gross-Pitaevskii equation due to the interaction of the condensate with collective excitations is 
considered. To do this the hydrodynamic approximation was used. In this approximation excitations are equivalent to
sound waves on a condensate background. This opens the possibility to use {\it methods
of quantum gravity and theory of quantum gauge fields} where the problem of calculation of effective action for gravitational and gauge backgrounds 
(or more precise, quantum corrections due to other quantum fields) is common problem. 

Many of the methods to approach the problem were developed during the last four decades. They are $\zeta$-function
regularization for determinants of operators, 
Schwinger-De-Witt-Seeley expansion of heat kernels \cite{Schwinger,DW,Seeley},
covariant methods of calculation of Seeley's coefficients and
the covariant perturbation technique for effective actions \cite{BV,BGVZ}.
Although the methods are well-known in field theory they are not so
familiar in condensed matter physics and the theory of coherent systems. 
An accurate account of corrections is very complicated even for a few first orders and  the solutions are obtained by using the covariant perturbation technique and curvature expansions.
 On the way effects nonlocal in space and time appeared. 
In quantum gravity framework such effects play significant role in the gravitational collapse problem, Hawking radiation and so on. 

However to benefit from it sometimes very general and 
only basic information about principle scales in systems in question is
required.
Indeed, this is the only information needed to extract leading logarithmic contributions while the calculation of other corrections is much more complicated. 
That is why we think that the developed approach can be widely used in many
problems which have not much to do with Bose-condensation of trapped atoms
or liquid Helium. 

At the end let us stop on other applications of the presented method.
It is easy to imagine other condensed matter examples of problems treatable by the same technique. Clearest and simplest of them are
antiferromagnets, Josephson arrays and superconductors in nonuniform (in space and time) external magnetic fields. However many of other physical systems
in a non-equilibrium background are potential field of applications.

\vspace{1cm}

\noindent
{\it Acknowledgment.}  
We are very grateful to Mike Gunn for frequent stimulating discussions and 
valuable comments.
We also want to thank Keith Burnnett, Dima Vasilevich, Martin Long, Ray Jones, 
Igor Lerner, Pavel Kornilovich and Ely Klepfish for the discussions of the problem. 
This work was supported by the Grant of Russian Fund of Fundamental
Investigations N 95-01-00548 and by the UK EPSRC Grants GR/L29156,
GR/K68356.

\vspace{1cm}

\newpage

\noindent
{\huge \bf Appendices}

\appendix

\def\theequation{\Alph{section}.\arabic{equation}}
\setcounter{equation}{0}

\section{Determinants of elliptic operators}

\setcounter{equation}{0}

Usually the first quantum correction to the effective action is ill defined. The point is it is divergent. This is just the well-known ultraviolet divergence of the
quantum field theory. Indeed, we can rewrite the functional determinant as
\begin{equation}
\ln\det A = {\rm Tr}\,\ln A =\ln \prod_n\limits \lambda_n 
=\sum\limits_n\ln\lambda_n
\label{20}
\end{equation}
where $\lambda_n$ are the eigenvalues of the operator $A$. This series is easy
to show to be divergent.

That means that one needs a regularization. This point was investigated very
thoroughly by many authors and it is found that in quantum gravity and gauge
theories the most appropriate regularizations are the analytical ones. The
functional determinants can be well defined in terms of the so called
$\zeta$-function.

At first let us use a  Wick rotation to produce
a Laplace operator from the wave operator. It allows to make use of methods for the 
evaluation of the determinants of elliptic operators \cite{Schwarz}. Now we are ready
to introduce the $\zeta$-function of an elliptic operator $A$:
$$
\zeta (s,A) = \sum_{i} \lambda_{i}^{-s}
$$
where $\{ \lambda_{i} \}$ are eigenvalues of the operator $A$. Then
$$
{\rm Tr}\,\ln A = - \left.\frac{\rm d}{{\rm d}s}\zeta(s,A)\right|_{s=0}
$$ 
To study the behavior of the $\zeta$-function it is common to use the formula
$$
 \lambda_{i}^{-s} =  \frac{1}{\Gamma(s)}\int\limits_0^\infty\! {\rm d}t\ 
t^{s-1} \exp(-t\lambda_{i}) 
\ .
$$
Summing over all nonzero eigenvalues of $A$ we get
\begin{equation}
\zeta (s,A) =  \frac{1}{\Gamma(s)}\int\limits_0^\infty\!{\rm d}t\ t^{s-1} 
( {\rm Tr}\, \exp(-tA) - L(A))
\label{b}
\end{equation}
where $L(A)$ is a number of zero-modes of $A$. 
Representation (\ref{b}) is valid when the integral converges.
For nonnegative operator it always does converge as $t \rightarrow +\infty$.
Conditions of the convergency of the integral as $t \rightarrow +0$ depend
on details of the operator $A$. For Laplace operator in 4D it converges as 
$t \rightarrow +0$ if ${\rm Re}\,s > 2$.

There exists the well-known Seeley expansion for the ${\rm Tr}\,(\exp(-tA))$:
$$
{\rm Tr}\,\exp(-tA) = \sum_{k\geq 0} \Phi_{-k}(A) t^{-k} + \rho(t) \ ,
$$
where $|\rho(t)|$ is bounded by a constant times $t$ as $t \rightarrow +0$.
$\Phi_{-k}(A)$ are called Seeley coefficients and play important role in the investigation of elliptic operators and their topological properties (for example in the Index theory). For example, for 4D Laplace operator in an external potential $E$ $\Phi_{-k} = 0$
$k\geq 3$, and
$$
\begin{array}{l}
\Phi_{-2} = (4\pi)^{-2} \ , \\
\Phi_{-1} = - (4\pi)^{-2}(E+\frac{1}{6} R) \ , \ 
\mbox{ where $R$ --- scalar curvature}\ ,
\end{array}
$$
and
$$
\Phi_{0} = (4\pi)^{-2} \left( -\frac{1}{30}\nabla^2 R + \frac{1}{72}R^2 -
\frac{1}{180}R_{\mu\nu} R^{\mu\nu}
 + \frac{1}{180}R_{\mu\nu\sigma\rho} R^{\mu\nu\sigma\rho} 
+\frac{1}{6}ER + \frac{1}{2}E^2 -\frac{1}{6}\nabla ^2 E\right) \ .
$$
Here $R_{\mu\nu\sigma\rho}$ and $R_{\mu\nu}$ 
are Riemann and Ricci tensors correspondingly. In the paper
we use the following definitions for them \cite{DNFW}:
$$
R = R^{\mu}_{\mu} \ , \qquad   R_{\mu\nu} = R^{\sigma}_{\mu\sigma\nu} \ ,
$$
and 
$$
R^{\lambda}_{\mu\nu\rho} = \frac{\partial \Gamma^{\lambda}_{\mu\nu}}{\partial x^{\rho}} - \frac{\partial \Gamma^{\lambda}_{\mu\rho}}{\partial x^{\nu}} +
\Gamma^{\sigma}_{\mu\nu}\Gamma^{\lambda}_{\rho\sigma} -
\Gamma^{\sigma}_{\mu\rho}\Gamma^{\lambda}_{\nu\sigma}  \ ,
$$
with the Christoffel symbols:
$$
\Gamma^{\sigma}_{\lambda\mu} = \frac{1}{2} g^{\nu\sigma}
( \frac{\partial g_{\mu\nu}}{\partial x^{\lambda}} + 
\frac{\partial g_{\lambda\nu}}{\partial x^{\mu}}  -
\frac{\partial g_{\mu\lambda}}{\partial x^{\nu}} )  \ .
$$
They are defined by a metric of a curved space. The latter coefficient $\Phi_0$ 
is particular important for the calculation of determinants.

Splitting (\ref{b}) into an integral over $[0,1]$ and one over $[1,+\infty]$ we get
\begin{eqnarray*}
\zeta(s) & = & \frac{1}{\Gamma(s)} \biggl( \sum_{k>0} 
\frac{\Phi_{-k}(A)}{s-k} +
\frac{\Phi_{0}(A)-L(A)}{s} + \\
& + & \int_1^{\infty}\!{\rm d}t\ {\rm Tr}\,\exp(-tA) t^{s-1}  + 
\int_0^{1}\!{\rm d}t\  \rho(t) t^{s-1} \biggr)
\end{eqnarray*}
The singularity at $s=0$ turns out to be removable since 
$\lim_{s\rightarrow 0} s\Gamma(s) = 1$ and $\zeta(s)$ is defined by analytical
continuation from the half-plane ${\rm Re}\, s > 0$.

From the relation $\ln\det A = -\zeta^{\prime}(0)$ we find that
\begin{eqnarray*}
\ln\det A & = &  \sum_{k>0} \frac{\Phi_{-k}(A)}{k} +
\Gamma^{\prime}(1)\Bigl( \Phi_{0}(A) - L(A) \Bigr) - 
 \int_1^{\infty}\!\frac{{\rm d}t}{t} {\rm Tr}\,\exp(-tA)  - \\  
 & - & \int_0^1\!\frac{{\rm d}t}{t} \left( {\rm Tr}\,\exp(-tA) 
- \sum_{k<0} \Phi_{-k}(A) t^{-k}
\right) \ .
\end{eqnarray*}
This is $\zeta$-regularized $\ln\det A$ which we use in the paper.

To conclude this section we give the relation between $\ln\det A$ and
$\ln\det \alpha A$ where $\alpha$ is a number parameter. It is easy to see that
$$
\zeta(s,\alpha A) = \zeta(s,A) \cdot \alpha^{-s}
$$
because $\lambda_i(\alpha A) = \alpha \lambda_i(A)$. This leads to the relation:
\begin{equation}
\ln\det \alpha A = \ln\det A + \ln\alpha\cdot\zeta(0,A) =
\ln\det A + \ln \alpha\cdot\Bigl(\Phi_0(A) - L(A) \Bigr)
\end{equation}
which we use intensively in the paper.

\section{Variational derivative of the effective action}
\setcounter{equation}{0}

In this appendix we give details of the calculation omitted in section 3
to do not interrupt the main stream of the consideration. To make the calculation
more efficient we make use covariant form of the quantum correction since
there exists well-know way to simplify covariant variation calculation
\cite{DNFW}.

We are interested in the following variational derivative
\begin{equation}
\frac{\delta }{\delta g^{\mu\nu}} \Phi_{0} (\Box )
\end{equation}
where
\begin{equation}
\Phi_{0} (\Box ) =
\frac{1}{(4\pi)^2} \int \!{\rm d}x\ \sqrt{-g}
\left(\frac{1}{72}R^2 -
\frac{1}{180}R_{\mu\nu} R^{\mu\nu}
 + \frac{1}{180}R_{\mu\nu\sigma\rho} R^{\mu\nu\sigma\rho}  + 
\frac{1}{6}RE +\frac{1}{2} E^2 \right) 
\end{equation}
For variation of the volume element we have
$$
\delta\sqrt{-g} = - \frac{1}{2\sqrt{-g}}\delta g_{\mu\nu}\Delta^{\mu\nu} =
\frac{\sqrt{-g}}{2}\delta g_{\mu\nu}g^{\mu\nu}
 = -\frac{\sqrt{-g}}{2}g_{\mu\nu}\delta g^{\mu\nu}
$$
where $\Delta^{\mu\nu}$ is defined by as
$$
\Delta^{\mu\nu} = g\cdot g^{\mu\nu}
$$
and the
following relation was used
$$
\delta g_{\mu\nu}g^{\nu\lambda} = -g_{\mu\nu}\delta g^{\nu\lambda}\ .
$$
For variations 
$R^2,R_{\mu\nu} R^{\mu\nu},R_{\mu\nu\sigma\rho} R^{\mu\nu\sigma\rho}$ 
one can obtain
\begin{eqnarray*}
{\delta} (RE) \!\!&=&\!\! \delta R\cdot E = \delta g^{12}R_{12} E 
+ g^{12}\delta R_{12}\cdot E \\
{\delta} R^2 \!\!&=&\!\! 2\delta R\cdot R = 2\delta g^{12}R_{12} R 
+  2 g^{12}\delta R_{12}\cdot R \\
{\delta} R_{12} R^{12} \!\!&=&\!\! \delta (g^{13}g^{24}R_{12}R_{34}) =
2\delta g^{12} R_{13} R_{2}^{\,\cdot\, 3} + 2 \delta R_{12} R^{12}
\\
{\delta} R_{1234} R^{1234} \!\!&=&\!\! \delta (g^{15}g^{26}g^{37}g^{48}R_{1234}R_{5678}) =
4\delta g^{12} R_{1345} R_{2}^{\,\cdot\, 345} + 2 \delta R_{1234} R^{1234}
\end{eqnarray*}
where numbers $1,2,3,\dots$ mean indices $\mu_1,\mu_2,\mu_3,\dots$ respectively.
Hence,
$$
\delta \Phi_0 = 
\frac{1}{(4\pi)^2}\int\!{\rm d}x\ \sqrt{-g}\,\delta g^{\mu\nu} 
\Biggl[
- \frac{1}{2}g_{\mu\nu}
\Biggl\{\frac{1}{72}R^2 - \frac{1}{180}R_{12}R^{12} + \frac{1}{180}
R_{1234}R^{1234}  + \frac{1}{6} RE +\frac{1}{2}E^2 \Biggr\} 
$$
$$
+ \frac{1}{36} R_{\mu\nu}R
 - \frac{1}{90}R_{\mu 1}R_k^{\,\cdot\,\nu}
 + \frac{1}{45}R_{\mu 123}R_\nu^{\,\cdot\, 123}
+ \frac{1}{6} R_{12} E\Biggr]
$$
$$
+ \frac{1}{(4\pi)^2}\int\!{\rm d}x\ \sqrt{-g} \Biggl[
\frac{1}{36}g^{12}\delta R_{12} R - \frac{1}{90} \delta R_{12} R^{12} + \frac{1}{90}
\delta R_{1234} R^{1234}  + \frac{1}{6} g^{12} \delta R_{12} E \Biggr]
$$
So we need to calculate last three terms. 

For variations of Ricci and Riemann tensors we have \cite{DNFW}:
\begin{eqnarray}
\delta R_{\mu\nu} &=& 
   \nabla_\nu(\delta \Gamma^1_{\mu 1})
 - \nabla_1(\delta \Gamma^1_{\mu\nu})
\nonumber\\ 
&=&
\frac{1}{2}g^{12}\left[ 
   \nabla_\mu\nabla_\nu\delta g_{12}
 - \nabla_1\nabla_\mu\delta g_{2\nu}
 - \nabla_1\nabla_\nu\delta g_{2\mu}
 + \nabla_1\nabla_2\delta g_{\mu\nu}
\right]
\label{varricci}\\
\delta R_{\mu\nu\sigma\rho} &=& 
 - g_{\mu1}\delta g^{12} R_{2\nu\sigma\rho}
 + g_{\mu1}\nabla_\rho(\delta \Gamma^1_{\nu\sigma})
 - g_{\mu1}\nabla_\sigma(\delta \Gamma^1_{\nu\rho})
\nonumber\\ 
&=&
 - g_{\mu1}\delta g^{12} R_{2\nu\sigma\rho}
 + \frac{1}{2}
\Bigl[
   \nabla_\rho\nabla_\sigma\delta g_{\mu\nu}
 + \nabla_\rho\nabla_\nu\delta g_{\mu\sigma}
 - \nabla_\rho\nabla_\mu\delta g_{\nu\sigma}
\nonumber\\ 
&&
 - \nabla_\sigma\nabla_\rho\delta g_{\mu\nu}
 - \nabla_\sigma\nabla_\nu\delta g_{\mu\rho}
 + \nabla_\sigma\nabla_\mu\delta g_{\nu\rho}
\Bigr]
\label{varriemann}
\end{eqnarray}
Let us consider covariant divergence of some vector $T^k$:
$$
\nabla_\mu T^\mu = \nabla^\mu T_\mu = \partial_\mu T^\mu 
+\Gamma_{\mu\nu}^\nu T^\mu = 
\frac{1}{\sqrt{-g}}\partial_\mu(\sqrt{-g}T^\mu)
$$
Hence
$$
\sqrt{-g}\nabla_\mu T^\mu = \partial_\mu (\sqrt{-g}T^\mu )
$$
is a pure divergence and can be dropped under integration.
Using this we can easily derive the following expression for variations
of the integrals we are interested in
\begin{eqnarray*}
&\!\!\!&\!\!\!\!\!\!
\int\!{\rm d}x\ \sqrt{-g} g^{\mu\nu}\delta R_{\mu\nu} R =
\int\!{\rm d}x\ \sqrt{-g} \delta g^{\mu\nu}
\left[ 
 + \frac{1}{2}\{\nabla_\mu,\nabla_\nu\}R
 - g_{\mu\nu}\Box R
\right]\\
&\!\!\!&\!\!\!\!\!\!
\int\!{\rm d}x\ \sqrt{-g} \delta R_{\mu\nu} R^{\mu\nu} = 
\frac{1}{2}\int\!{\rm d}x\ \sqrt{-g} \delta g^{\mu\nu}
\left[
 - \Box R_{\mu\nu}
 - g_{\mu\nu}\nabla_1\nabla_2 R^{12}
 + \nabla_1\nabla_\mu R_\nu^{\,\cdot\,1}
 + \nabla_1\nabla_\nu R_\mu^{\,\cdot\,1}
\right]\\
&\!\!\!&\!\!\!\!\!\!
\int\!{\rm d}x\ \sqrt{-g} \delta R_{\mu\nu\sigma\rho} R^{\mu\nu\sigma\rho} = 
\int\!{\rm d}x\ \sqrt{-g} \delta g^{\mu\nu}
\left[
 - R_{\mu123} R_\nu^{\,\cdot\,123}
 + 2\nabla^2\nabla^1 R_{\mu12\nu}
\right] \\
&\!\!\!&\!\!\!\!\!\!
\int\!{\rm d}x\ \sqrt{-g} g^{\mu\nu}\delta R_{\mu\nu} E =
\int\!{\rm d}x\ \sqrt{-g} \delta g^{\mu\nu}
\left[ 
 + \frac{1}{2}\{\nabla_\mu,\nabla_\nu\}E
 - g_{\mu\nu}\Box E
\right]
\end{eqnarray*}
Summing all terms we obtain
\begin{eqnarray*}
\frac{\delta }{\delta g^{\mu\nu}} \Phi_{0} 
\!\!& =\!\! &
 \frac{1}{(4\pi)^2}
\sqrt{-g} 
\Biggl\{
 - \frac{1}{2} g_{\mu\nu} 
\Bigl[ 
   \frac{1}{72}R^2
 - \frac{1}{180} R_{\sigma\rho}R^{\sigma\rho} 
 + \frac{1}{180}R_{\sigma\rho\alpha\beta}R^{\sigma\rho\alpha\beta}
+ \frac{1}{6} RE + \frac{1}{2} E^2
\Bigr] 
\\
\!\!& +\!\! & 
\frac{1}{6} R_{\mu\nu}E
  + \frac{1}{36} R_{\mu\nu}R
 - \frac{1}{90}R_{\mu\sigma}R_{\nu}^{\,\cdot\,\sigma} 
 + \frac{1}{90}R_{\mu\sigma\rho\alpha}R_{\nu}^{\,\cdot\,\sigma\rho\alpha}
 - \frac{1}{36}g_{\mu\nu} \Box R
\\
\!\!& +\!\! & 
\frac{1}{72} \{\nabla _{\mu},\nabla_{\nu}\} R  
 - \frac{1}{6}g_{\mu\nu} \Box E 
 + \frac{1}{12} \{\nabla _{\mu},\nabla_{\nu}\} E
\\
\!\!& +\!\! & 
\frac{1}{180} 
\Bigl[ 
 - 2\nabla^{\sigma}\nabla_{\nu} R_{\mu\sigma}
 + \Box R_{\mu\nu}
 + g_{\mu\nu}\nabla^{\sigma}\nabla^{\rho} R_{\sigma\rho} 
\Bigr]
 + \frac{1}{45} 
   \nabla^{\rho}\nabla^{\sigma}R_{\mu\sigma\rho\nu} 
\Biggr\}
\end{eqnarray*}



\begin{thebibliography}{99}
\bibitem{Osc}The $s$-wave scattering length can be estimated as
$l \sim 100 a_{0}$, where
$a_{0}$ is the Bohr radius (i.e., $l \sim 5 \cdot 10^{-7} \mbox{cm}$);
\bibitem{note2}
Using field-theoretical terminology the analysis by
Wu and Griffin \cite{WG} corresponds to a classical level while
term (\ref{Gamma1}) is a first quantum correction. Since paper
\cite{WG} has a title ``Quantized hydrodynamic model..." a confusion
can occur. In this way, the only self-consistent possibility to
introduce non-condensed fraction in the
consideration is to study the action (\ref{ac0}) with
(at least 1-loop) quantum corrections.
\bibitem{note3}
It is not absolutely true
for two-point correlation function only when the approach gives an partial derivatives differential equation which still has to be solved; it can be done
numerically and is not principal problem.
\end{thebibliography}
\end{document}